\pdfoutput=1
\RequirePackage{fix-cm} % exact Computer Modern sizes (title 20pt, figure-schema 6.5/7.5pt)
\documentclass[11pt]{article}

\usepackage{float}
\usepackage{placeins}

\usepackage{graphicx}
\usepackage{caption}
\usepackage{subfig} % \subfloat, as in main.tex
\usepackage{amsmath}
\usepackage{amssymb}

\usepackage{natbib}
\setcitestyle{numbers,round,comma}
\usepackage{geometry}
\usepackage{listings}
\usepackage{booktabs}
\usepackage{tabularx}
\usepackage{multirow}
\usepackage{adjustbox}
\usepackage{paralist}
\usepackage{tikz}
\expandafter\ifx\csname iflatexml\endcsname\relax\newif\iflatexml\latexmlfalse\fi
\usepackage[most]{tcolorbox}
\usetikzlibrary{positioning, calc, fit, backgrounds, arrows.meta}

\usepackage{xcolor}
\usepackage{colortbl}
\usepackage{xspace}
\usepackage{etoolbox}
\definecolor{eegblue}{HTML}{006CA3}
\definecolor{eegorange}{HTML}{F7941D}
\definecolor{eeglight}{HTML}{E8F4FD}
\definecolor{eeggray}{HTML}{64748B}

\makeatletter
\providecommand{\nrlabel}[2]{\phantomsection\def\@currentlabelname{#1}\label{#2}\nobreak}
\makeatother

\definecolor{codebg}{HTML}{F8FAFC}
\definecolor{codeframe}{HTML}{D5DDE7}
\definecolor{codekw}{HTML}{005C8B}
\definecolor{codestr}{HTML}{A16013}
\definecolor{codecomment}{HTML}{64748B}
\definecolor{codebuiltin}{HTML}{0F172A}
\makeatletter\def\zifour@opt{4}\def\zifour@altopt{6}\makeatother % code font: Inconsolata (zi4) with upright quotes
\lstdefinestyle{eegpython}{
    language=Python,
    basicstyle=\fontencoding{T1}\ttfamily\small,
    columns=fullflexible,
    keepspaces=true,
    backgroundcolor=\color{codebg},
    frame=single,
    framerule=0.4pt,
    rulecolor=\color{codeframe},
    framesep=5pt,
    framexleftmargin=0pt,
    framexrightmargin=0pt,
    xleftmargin=5.4pt,
    xrightmargin=5.4pt,
    breaklines=true,
    breakatwhitespace=true,
    keywordstyle=\color{codekw},
    stringstyle=\color{codestr},
    commentstyle=\color{codecomment},
    emphstyle=\color{codebuiltin}\bfseries,
    emph={EEGDashDataset,EEGDash,DataLoader,create_fixed_length_windows,DS001849},
    showstringspaces=false,
    numbers=none,
    tabsize=4,
    aboveskip=6pt,
    belowskip=8pt,
}
\usepackage{titlesec}
\usepackage{fancyhdr}
\usepackage{authblk}

\titleformat{\section}{\Large\bfseries}{\textcolor{eegblue}{\thesection}}{1em}{\textcolor{eegblue}}[\nobreak\vspace{2pt}\nobreak\titleline{\color{eegorange}\leaders\hrule height 1.2pt\hfill\kern0pt}\nobreak]
\titleformat{\subsection}{\large\bfseries}{\textcolor{eegblue!85!black}{\thesubsection}}{1em}{\textcolor{eegblue!85!black}}

\fancypagestyle{plain}{\fancyhf{}\fancyfoot[C]{\small\color{eeggray}\thepage}}

\newcommand{\eegdash}{EEGDash\xspace}
\newcommand\blfootnote[1]{\begingroup\renewcommand\thefootnote{}\footnotetext{#1}\endgroup}
\newcommand{\upd}[1]{#1}
\newcommand{\Ndatasets}{\upd{892}}
\newcommand{\Nsubjects}{\upd{43\,911}}
\newcommand{\Nhours}{\upd{126\,672}}
\newcommand{\Nrecords}{\upd{273\,123}}
\newcommand{\Nsize}{\upd{71.7}}
\newcommand{\Nsnapshot}{10 October 2026}
\newcommand{\Nversion}{0.9.1}

\newcommand{\figalt}[1]{}

\graphicspath{{figures/}{./}}

\usepackage{hyperref}
\hypersetup{colorlinks=true, linkcolor=eegblue, citecolor=eegblue, urlcolor=eegblue,
  pdftitle={EEGDash: An open-source platform for machine learning on public neurophysiological data},
  pdfauthor={Bruno Aristimunha, Aviv Dotan, Pierre Guetschel, Aman Jaiswal, Gal Ashkenazi, Dung Truong, Kuntal Kokate, Amit Majumdar, Oren Shriki, Arnaud Delorme}}
\usepackage{orcidlink} % ORCID iDs next to author names, as in the Database version

\begin{document}

\title{EEGDash: An open-source platform for machine learning on public neurophysiological data}

% Marks as in the Database version: $\ast$ corresponding author, $\dagger$ equal contribution.
\author[1,2,3,$\ast$,$\dagger$]{Bruno Aristimunha\,\orcidlink{0000-0001-5258-2995}}
\author[4,$\dagger$]{Aviv Dotan\,\orcidlink{0000-0003-4515-9333}}
\author[5]{Pierre Guetschel\,\orcidlink{0000-0002-8933-7640}}
\author[1]{Aman Jaiswal}
\author[4]{Gal Ashkenazi}
\author[1]{Dung Truong\,\orcidlink{0000-0003-4540-3551}}
\author[1]{Kuntal Kokate}
\author[1]{Amit Majumdar\,\orcidlink{0000-0002-0860-6686}}
\author[4,$\dagger$]{Oren Shriki\,\orcidlink{0000-0003-1129-4799}}
\author[1,6,$\ast$,$\dagger$]{Arnaud Delorme\,\orcidlink{0000-0002-0799-3557}}

\affil[1]{University of California San Diego, La Jolla, CA, United States}
\affil[2]{Yneuro, Paris, France}
\affil[3]{Universit\'{e} Paris-Saclay, Inria, Palaiseau, France}
\affil[4]{Department of Cognitive and Brain Sciences, Ben-Gurion University of the Negev, Beer-Sheva, Israel}
\affil[5]{Donders Institute for Brain, Cognition and Behaviour, Radboud University, Nijmegen, The Netherlands}
\affil[6]{Centre de Recherche Cerveau et Cognition (CerCo), CNRS UMR 5549, Universit\'{e} Toulouse III -- Paul Sabatier, Toulouse, France}

\date{}

\renewcommand\Authfont{\normalfont\bfseries\normalsize}
\renewcommand\Affilfont{\normalfont\small\itshape\color{eeggray}}
\renewcommand\Authands{ and }
\setlength{\affilsep}{0.7em}
\makeatletter
\renewcommand{\@maketitle}{%
  \null\vskip 0.5em%
  \begin{center}%
    \includegraphics[width=0.46\textwidth]{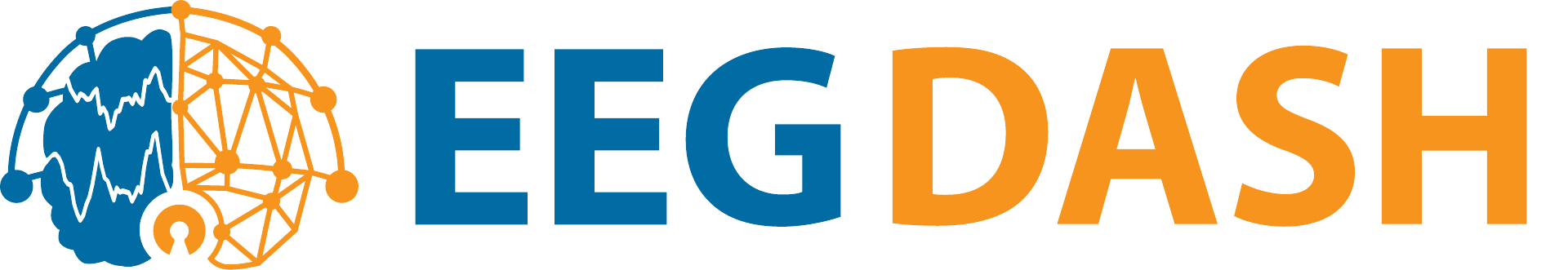}\\[1.1em]%
    {\fontsize{20}{24}\selectfont\bfseries\color{eegblue}\@title\par}%
    \vspace{0.4em}%
    {\color{eegorange}\rule{0.86\textwidth}{1.1pt}}\\[1.0em]%
    {\@author}\par%
  \end{center}%
  \vskip 0.6em%
}
\makeatother
\renewenvironment{abstract}{%
  \begin{tcolorbox}[enhanced,colback=eeglight,colframe=eegblue!55!white,
      boxrule=0.6pt,arc=2pt,left=10pt,right=10pt,top=7pt,bottom=7pt]%
  {\color{eegblue}\bfseries Abstract}\par\smallskip\small\noindent\ignorespaces
}{%
  \end{tcolorbox}%
}

\maketitle
\blfootnote{\textsuperscript{$\ast$}Corresponding authors. \href{mailto:baristimunha@ucsd.edu}{baristimunha@ucsd.edu} (B.A.); \href{mailto:adelorme@ucsd.edu}{adelorme@ucsd.edu} (A.De.)}
\blfootnote{\textsuperscript{$\dagger$}B.A., A.Do., O.S.\ and A.De.\ contributed equally.}

\begin{abstract}
Public recordings of electroencephalography (EEG), magnetoencephalography (MEG), intracranial EEG (iEEG), functional near-infrared spectroscopy (fNIRS) and electromyography (EMG) are easy to find and hard to train on. Turning a deposit into model input still means writing download, parsing, repair and windowing code for each dataset, and a dataset that meets a metadata standard can still fail to load. \eegdash{} is an open-source registry over public Brain Imaging Data Structure (BIDS) datasets from the OpenNeuro and NEMAR archives and from local deposits. Each dataset becomes a Python class; format defects are repaired lazily; signal handling, windowing and format checking are delegated to MNE-Python, Braindecode and the official BIDS validator. Semantic search, a web browser, publication-derived dataset tags and a feature-extraction module share the registry. \eegdash{} indexes the metadata of \Ndatasets{} public BIDS datasets (\Nsubjects{} participants and \Nhours{} hours across the five modalities), loads any of them into a PyTorch dataset without custom data-loading code, and shows on 503 OpenNeuro EEG datasets that passing the BIDS validator does not predict whether a dataset loads (Spearman $\rho = -0.05$), so an archive built for machine learning has to track both.\\[0.5em]
\textbf{Database URL:} \url{https://eegdash.org}
\end{abstract}


%% ===== BODY =====
\section*{Introduction}\nrlabel{Introduction}{sec:background}

Public archives of neurophysiological recordings have grown quickly under the Brain Imaging Data Structure (BIDS) \citep{gorgolewski2016bids, pernet2019eegbids}: OpenNeuro and NEMAR alone hold hundreds of EEG, MEG, iEEG, fNIRS and EMG datasets \citep{markiewicz2021openneuro, delorme2022nemar}. BIDS fixes the layout of a deposit and the metadata files that travel with it, and the official BIDS validator checks that layout. The FAIR principles describe what such data should be: findable, accessible, interoperable and reusable \citep{wilkinson2016fair}. The first two are largely met. The second two break where the data meet a model. Cross-dataset models cannot train on data they cannot load \citep{kostas2021bendr, yang2023biot, jiang2024labram, xiao2025brainomni}, and a deposit becomes model input only after download, parsing, repair and windowing code has been written for it \citep{sculley2015debt}.

\begin{figure*}[tbp]
\centering
\iflatexml
\includegraphics[width=0.422\textwidth]{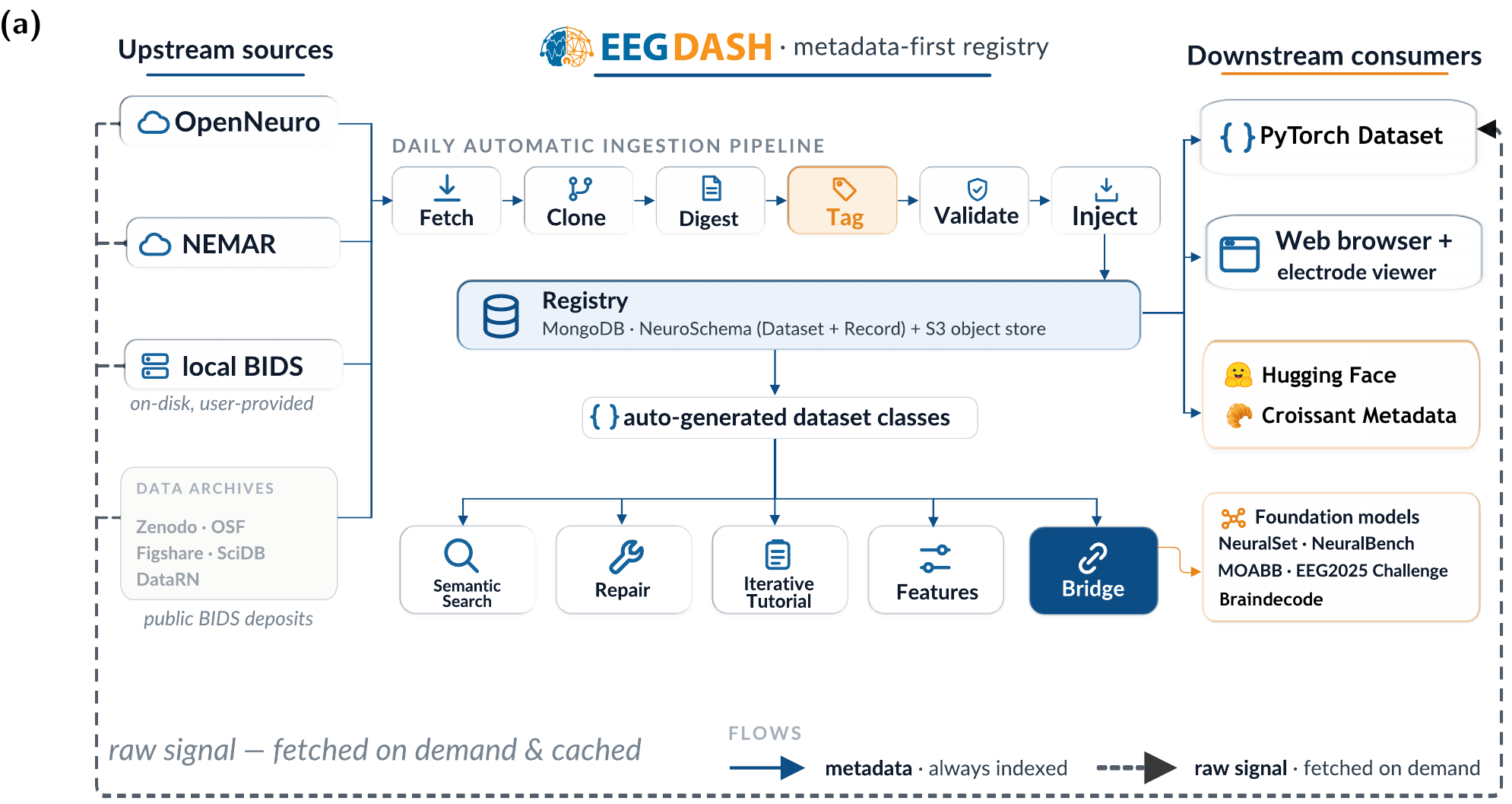}%
\includegraphics[width=0.558\textwidth]{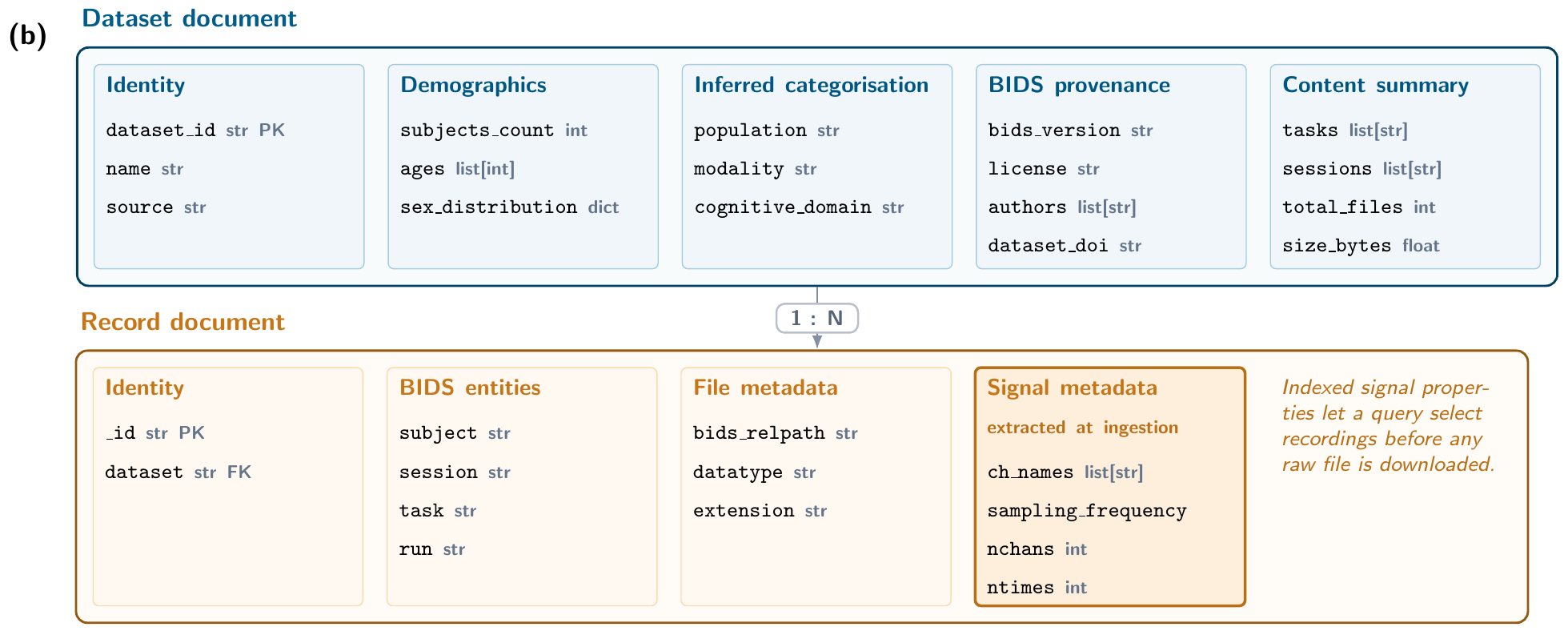}
\else
\begin{tikzpicture}
  \node[anchor=north west, inner sep=0] (panelA) at (0,0) {\includegraphics[width=0.9825\textwidth]{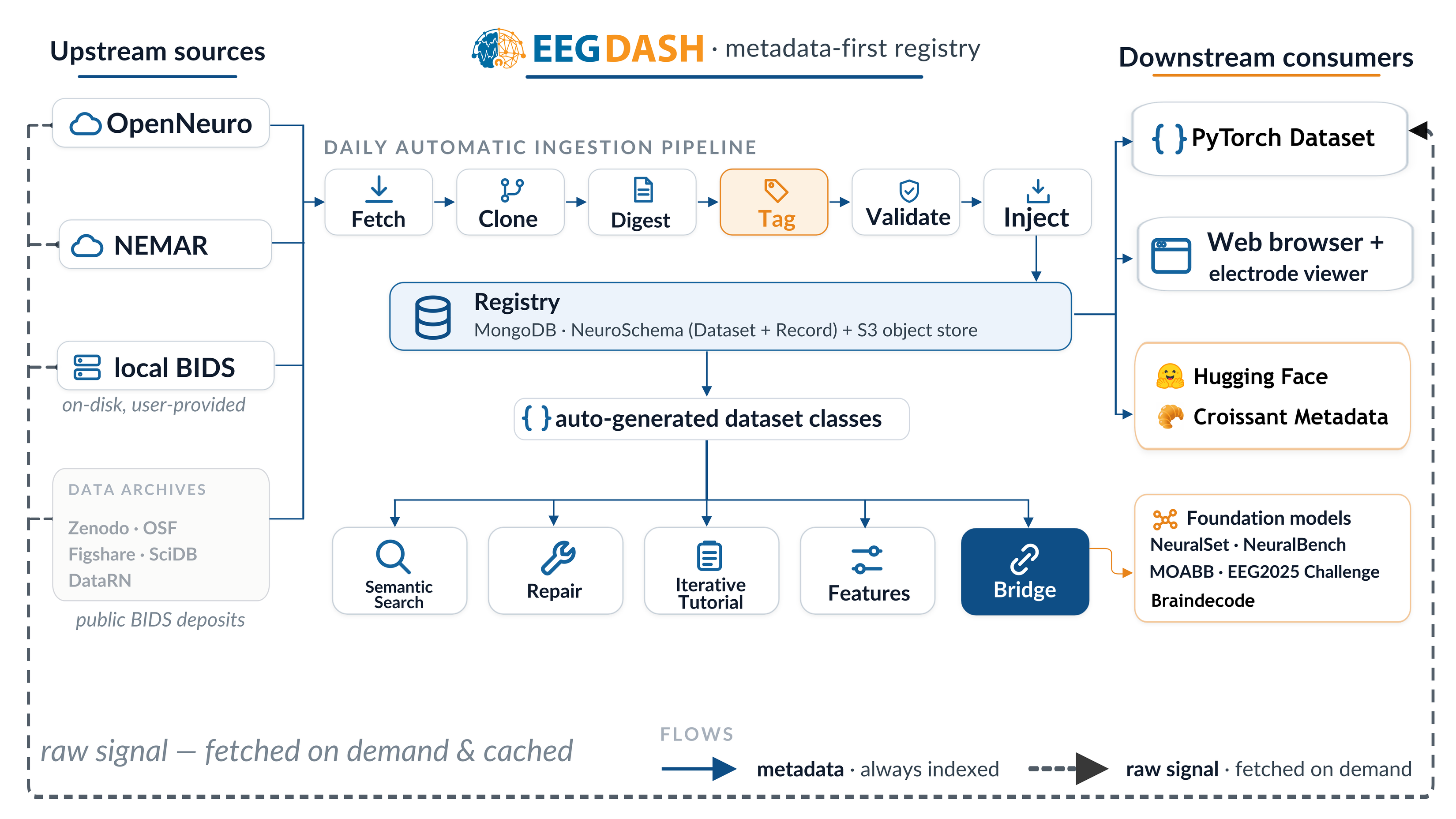}};
  \node[overlay, anchor=north east, font=\sffamily\bfseries\fontsize{10}{12}\selectfont] at ([xshift=-4pt]panelA.north west) {(a)};
  \pgfresetboundingbox\useasboundingbox (-14.6pt,0) rectangle ({\textwidth-14.6pt},{-0.9825*0.5625*\textwidth}); % exactly \textwidth wide: the (a) border sits on the (b) frame
\end{tikzpicture}\\[5pt]
\begin{adjustbox}{max width=\textwidth}
\begin{tikzpicture}[node distance=3mm and 3mm]
\def\ttype#1{{\color{eeggray}\sffamily\fontsize{6.5}{8}\selectfont\bfseries #1}}
\def\tfield#1{\texttt{\fontsize{7.5}{9}\selectfont #1}}
\def\tgrp#1#2{{\sffamily\fontsize{8}{9.5}\selectfont\bfseries\color{#1!80!black}#2}}
\tikzset{
  card/.style={rectangle, text width=3.05cm, inner sep=4pt, align=left},
  dscard/.style={card},
  reccard/.style={card},
  sigcard/.style={card},
  cardbg/.style={rectangle, rounded corners=2pt, line width=0.4pt, inner sep=0pt},
  frame/.style={rectangle, rounded corners=4pt, line width=0.8pt, inner sep=6pt},
  dsframe/.style={frame, draw=eegblue!60!black, fill=eegblue!3},
  recframe/.style={frame, draw=eegorange!60!black, fill=eegorange!3},
  relchip/.style={rectangle, rounded corners=3pt, fill=white, draw=eeggray!45, inner xsep=5pt, inner ysep=2.5pt,
                  font=\sffamily\bfseries\fontsize{7.5}{9}\selectfont, text=eeggray!85!black},
}
% ---- dataset document (one card per field group) ----
\node[dscard] (d1) at (0,0)
  {\tgrp{eegblue}{Identity}\\[2pt]
   \tfield{dataset\_id} \ttype{str PK}\\
   \tfield{name} \ttype{str}\\
   \tfield{source} \ttype{str}};
\node[dscard, right=of d1.north east, anchor=north west] (d2)
  {\tgrp{eegblue}{Demographics}\\[2pt]
   \tfield{subjects\_count} \ttype{int}\\
   \tfield{ages} \ttype{list[int]}\\
   \tfield{sex\_distribution} \ttype{dict}};
\node[dscard, right=of d2.north east, anchor=north west] (d3)
  {\tgrp{eegblue}{Inferred categorisation}\\[2pt]
   \tfield{population} \ttype{str}\\
   \tfield{modality} \ttype{str}\\
   \tfield{cognitive\_domain} \ttype{str}};
\node[dscard, right=of d3.north east, anchor=north west] (d4)
  {\tgrp{eegblue}{BIDS provenance}\\[2pt]
   \tfield{bids\_version} \ttype{str}\\
   \tfield{license} \ttype{str}\\
   \tfield{authors} \ttype{list[str]}\\
   \tfield{dataset\_doi} \ttype{str}};
\node[dscard, right=of d4.north east, anchor=north west] (d5)
  {\tgrp{eegblue}{Content summary}\\[2pt]
   \tfield{tasks} \ttype{list[str]}\\
   \tfield{sessions} \ttype{list[str]}\\
   \tfield{total\_files} \ttype{int}\\
   \tfield{size\_bytes} \ttype{float}};
\begin{scope}[on background layer]
\node[dsframe, fit=(d1)(d2)(d3)(d4)(d5)] (dsframe) {};
\end{scope}
\begin{scope}[on background layer]
\foreach \n in {d1,d2,d3,d4,d5}
  \node[cardbg, draw=eegblue!35, fill=eegblue!6, fit=(\n.north west)(\n.east |- d5.south)] {};
\end{scope}
\node[anchor=south west, font=\sffamily\bfseries\fontsize{9}{11}\selectfont, text=eegblue!85!black]
      at ([xshift=-2pt,yshift=3pt]dsframe.north west) {Dataset document};
% ---- record document ----
\node[reccard, below=10mm of dsframe.south west, xshift=6pt, anchor=north west] (r1)
  {\tgrp{eegorange}{Identity}\\[2pt]
   \tfield{\_id} \ttype{str PK}\\
   \tfield{dataset} \ttype{str FK}};
\node[reccard, right=of r1.north east, anchor=north west] (r2)
  {\tgrp{eegorange}{BIDS entities}\\[2pt]
   \tfield{subject} \ttype{str}\\
   \tfield{session} \ttype{str}\\
   \tfield{task} \ttype{str}\\
   \tfield{run} \ttype{str}};
\node[reccard, right=of r2.north east, anchor=north west] (r3)
  {\tgrp{eegorange}{File metadata}\\[2pt]
   \tfield{bids\_relpath} \ttype{str}\\
   \tfield{datatype} \ttype{str}\\
   \tfield{extension} \ttype{str}};
\node[sigcard, right=of r3.north east, anchor=north west] (r4)
  {\tgrp{eegorange}{Signal metadata}\\
   {\sffamily\fontsize{6.5}{8}\selectfont\bfseries\color{eegorange!75!black}extracted at ingestion}\\[2pt]
   \tfield{ch\_names} \ttype{list[str]}\\
   \tfield{sampling\_frequency}\\
   \tfield{nchans} \ttype{int}\\
   \tfield{ntimes} \ttype{int}};
\node[right=of r4.north east, anchor=north west, text width=2.7cm, inner sep=4pt, align=left,
      font=\sffamily\itshape\fontsize{7.5}{9}\selectfont, text=eegorange!75!black] (r5)
  {Indexed signal properties let a query select recordings before any raw file is downloaded.};
\begin{scope}[on background layer]
\node[recframe, fit=(r1)(r2)(r3)(r4)(r5)] (recframe) {};
\end{scope}
\begin{scope}[on background layer]
\foreach \n in {r1,r2,r3}
  \node[cardbg, draw=eegorange!35, fill=eegorange!6, fit=(\n.north west)(\n.east |- r4.south)] {};
\node[cardbg, draw=eegorange!75!black, fill=eegorange!15, line width=1.0pt, fit=(r4)] {};
\end{scope}
\node[anchor=south west, font=\sffamily\bfseries\fontsize{9}{11}\selectfont, text=eegorange!80!black]
      at ([xshift=-2pt,yshift=3pt]recframe.north west) {Record document};
% ---- relation ----
\draw[eeggray!80, line width=0.7pt, -{Latex[length=5pt]}] (dsframe.south -| d3.south) -- (recframe.north -| d3.south)
      node[midway, relchip] {1 : N};
\node[anchor=north east, font=\sffamily\bfseries\fontsize{10}{12}\selectfont] at ([xshift=-6pt, yshift=12pt]dsframe.north west) {(b)};
\end{tikzpicture}
\end{adjustbox}
\fi
\caption{\textbf{\eegdash{} end to end.} (a)~A daily pipeline ingests upstream BIDS archives (left) into a metadata-first registry that exposes \eegdash{}'s own surfaces and serves downstream consumers (right); indexed metadata flows through the core (solid), the raw signal is fetched on demand and cached (dashed). (b)~The registry schema: a dataset document holds study-level metadata and links to $N$ record documents, which hold the BIDS entities, the file location and the signal properties indexed at ingestion. Representative fields; the full schema ships with the Python package.}
\figalt{Panel a: flow diagram showing OpenNeuro, NEMAR, local BIDS and other public archives feeding an EEGDash metadata registry that powers semantic search, repair, tutorials, feature extraction, bridges, the web browser, PyTorch datasets, Hugging Face Croissant metadata, foundation models, NeuralSet, NeuralBench, MOABB, the EEG2025 challenge and Braindecode. Panel b: schema diagram with a dataset document (identity, demographics, inferred categorisation, BIDS provenance, content summary) linked one-to-many to record documents (identity, BIDS entities, file metadata, signal metadata).}
\label{fig:ecosystem}
\end{figure*}
The tools that could close this distance come in two kinds, and neither closes it alone. Archives host data at scale but stop at storage: OpenNeuro, NEMAR and DANDI (\url{https://dandiarchive.org}) \citep{markiewicz2021openneuro, delorme2022nemar} deliver a dataset as raw files that a researcher must parse, repair and window before a model sees it. Curated machine-learning libraries make the opposite trade: MOABB \citep{chevallier2024moabb}, Brainsets and TorchEEG \citep{zhang2024torcheeg} expose PyTorch-ready loaders over a hand-picked set of datasets, so breadth is what they give up. Between the two sits a drift that nobody measures. The loader that reads a deposit and the standard the deposit follows are both living software, maintained with little long-term funding \citep{westner2025cycling}, and the earlier survey of neuroelectrophysiology archives recorded formats and metadata but not whether a deposit loads \citep{subash2023comparison}. Two things have not been measured on a whole archive: whether a public BIDS archive can be trained on without per-dataset engineering, and whether the BIDS validator is the right gate for deciding that. In the OpenNeuro EEG corpus, one dataset in three passes the BIDS validator; the libraries that do load cleanly are curated by hand and cover a small part of what the archives hold.

\eegdash{} is built on one principle, metadata-first lazy access: the registry indexes the metadata of every recording, and a researcher downloads only the signal a model consumes \citep{mohan2021datastalls, wagner2022fairlybig}. Indexed metadata flows through the \eegdash{} core; the raw signal bypasses it and arrives on demand (Figure~\ref{fig:ecosystem}). The same principle decides what \eegdash{} owns and what it delegates. Signal processing runs through MNE-Python \citep{gramfort2013mne}, BIDS parsing through MNE-BIDS \citep{appelhoff2019mnebids}, windowing and training through Braindecode and PyTorch \citep{braindecode, paszke2019pytorch}, validation through the official BIDS validator, and HPC dispatch through the Neuroscience Gateway \citep{sivagnanam2013nsg}; \eegdash{} re-implements none of them and \href{https://github.com/mne-tools/mne-bids/issues?q=is%3Apr%20author%3A%40bruAristimunha}{sends generalisable fixes upstream}. It owns the registry, the repair layer and the dataset classes, the parts that have no maintained equivalent, because a private fork would drift from upstream as the archive drifts from BIDS. Repair is lazy: a defect is detected when a recording is first read and corrected in memory, so the deposit itself is never rewritten.

The registry, the web browser and the Python package are public at \url{https://eegdash.org} and need no login. This paper describes \eegdash{}, which (1) indexes signal metadata once at ingestion, so recordings are selected before any download; (2) loads every catalogued BIDS recording, repaired on the fly, into a PyTorch dataset through one class, as windowed signal or as a table of features; and (3) measures BIDS-validator compliance and loader success on the same OpenNeuro EEG datasets and finds that one does not predict the other.

\section*{Materials and Methods}\nrlabel{Materials and Methods}{sec:methods}

\subsection*{The registry and its query surface}\nrlabel{The EEGDash database}{sec:database}\nrlabel{Metadata Schema}{sec:schema}

The registry is a MongoDB database with two collections (Figure~\ref{fig:ecosystem}b). A \textbf{dataset document} holds the study-level metadata used for discovery: identity and source, demographics, the inferred population, modality and cognitive-domain tags, BIDS provenance and licence, and a content summary. It links to \emph{N} \textbf{record documents}, one per recording, with the BIDS entities, the storage location, and the signal properties read from the file header at ingestion: channel names, channel count, sampling frequency and duration. Both collections are queryable before any download: through an open REST API at \href{https://data.eegdash.org}{data.eegdash.org} that accepts MongoDB-style filters, through the Python package, and by semantic search over the dataset documents. Because the signal properties are indexed, a query can narrow the archive to a few studies and then to specific recordings (for example, 64-channel recordings sampled at 256~Hz or higher) without touching a signal file. From the populated registry, \eegdash{} generates one Python class per dataset. Every class inherits from \texttt{EEGDashDataset}, shares one constructor signature and returns a PyTorch \texttt{Dataset} built on Braindecode; a continuous-integration test checks this contract on every release.

\subsection*{Web browser and electrode viewer}\nrlabel{Dataset Browser}{sec:browser}

\begin{figure*}[!t]
    \centering
    \subfloat[Dataset page and field card.\label{fig:browser-page}]{%
        \includegraphics[width=0.426\textwidth]{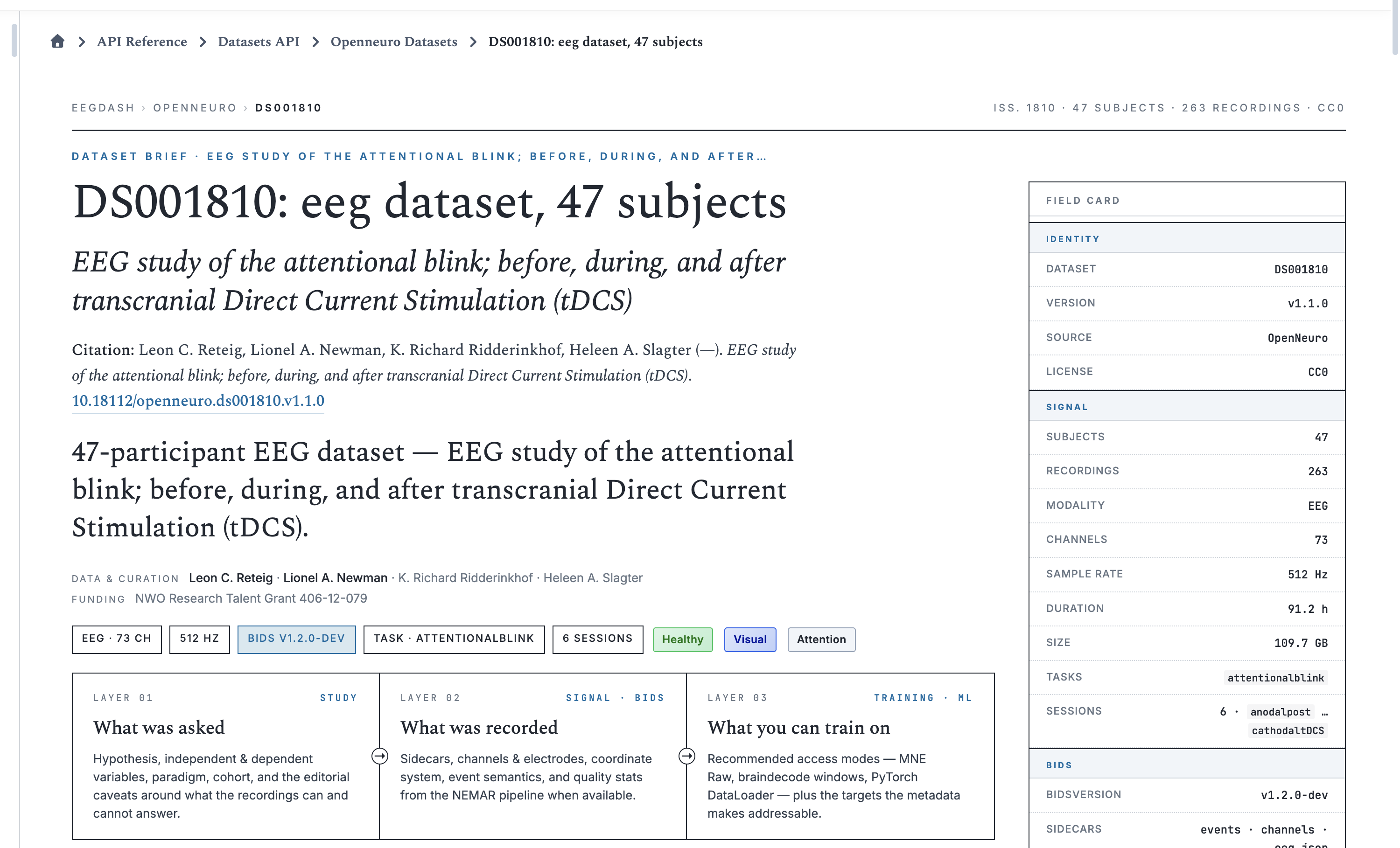}}
    \hfill
    \subfloat[\textsc{Signal\,$\cdot$\,Electrodes} viewer.\label{fig:browser-viewer}]{%
        \includegraphics[width=0.554\textwidth]{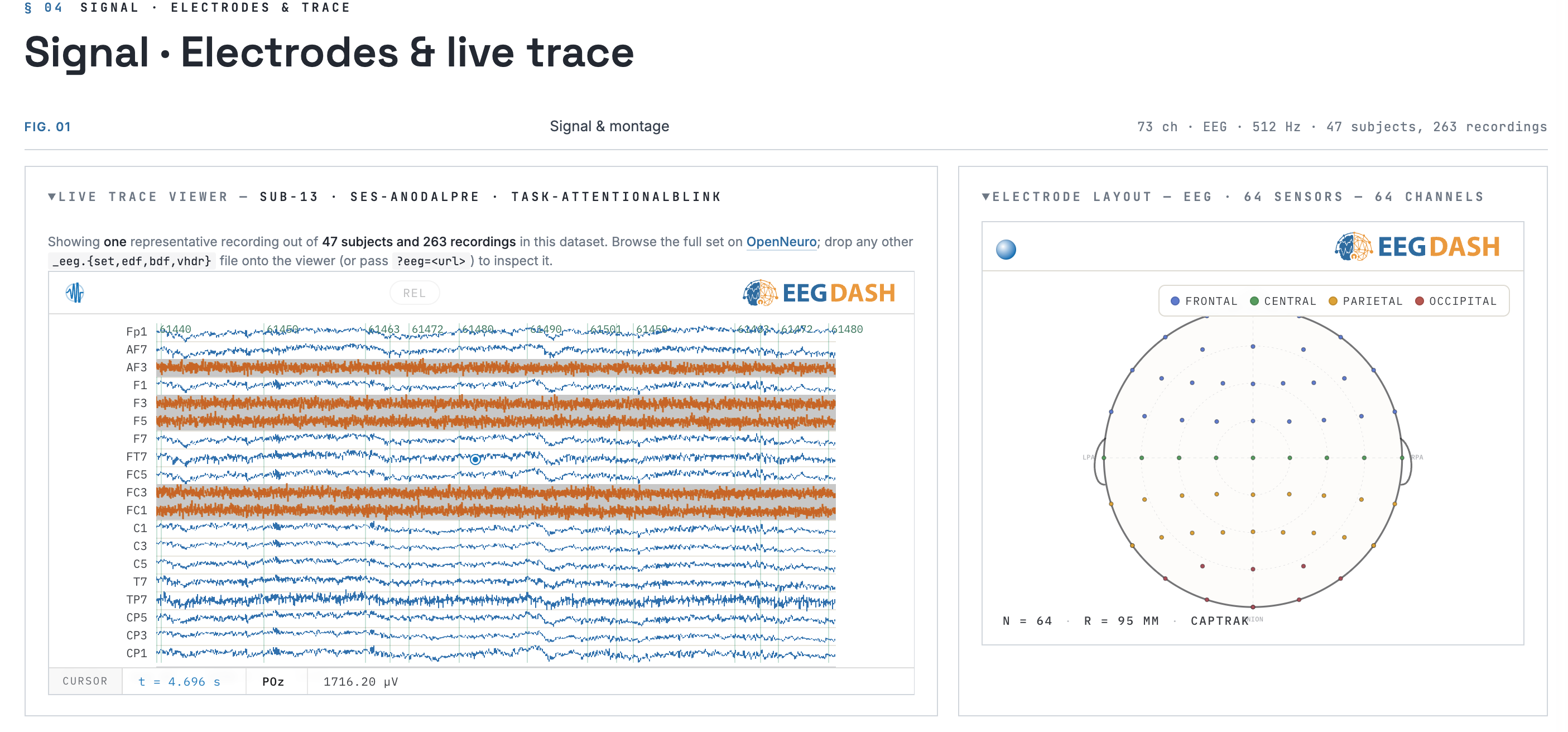}}
    \caption{The \eegdash{} web browser (\url{https://eegdash.org}). (a)~A dataset page: the registry metadata as a field card, a paste-ready code snippet and an ML-readiness summary. (b)~The signal and electrodes viewer: a multichannel trace and the scalp montage (here 64~EEG sensors, coloured by region) rendered from the electrodes sidecars, so channel coverage is visible before any download.}
    \figalt{Two screenshots of the EEGDash web interface. The first shows a dataset page with metadata cards and a code snippet. The second shows an electrode and signal viewer with multichannel EEG traces beside a scalp topography plot.}
    \label{fig:browser}
\end{figure*}

A web interface at \url{https://eegdash.org} presents the registry as a sortable, filterable catalogue backed by the same query primitives, so a filter built in the browser can be copied into a notebook. Each dataset page shows a field card, a paste-ready code snippet, NEMAR processing statistics, and an electrode viewer that renders the scalp montage from the \texttt{electrodes} sidecars, so channel coverage is visible before any download (Figure~\ref{fig:browser}).

\subsection*{Sources and ingestion}\nrlabel{The ingestion pipeline}{sec:pipeline}\nrlabel{Data Collection}{sec:collection}\nrlabel{Ingestion and Registry}{sec:ingestion}

Datasets come from OpenNeuro \citep{markiewicz2021openneuro} and NEMAR \citep{delorme2022nemar}, and from BIDS deposits on Zenodo, OSF, Figshare, SciDB and DataRN. Dataset structure follows BIDS \citep{gorgolewski2016bids}, read through MNE-BIDS \citep{appelhoff2019mnebids}; HED event tags \citep{robbins2021hed} are carried over where a source supplies them. A daily, metadata-only pipeline polls each archive's public API and runs five stages per dataset. \emph{Fetch} writes a per-dataset manifest. \emph{Clone} materialises the files over the filesystem, git-annex or S3. \emph{Digest} parses the BIDS layout and the signal headers into dataset- and record-level metadata and the electrode montages. \emph{Validate} checks the digest against a Pydantic schema (types, value ranges, cross-field consistency) and checks that the derived ML-ready format reproduces the source signal numerically and loads in PyTorch and scikit-learn. \emph{Inject} writes only the create, update or skip difference, so a refresh is idempotent and runs on small runners. Between digest and validate, a tagger built on a large language model assigns a population, a modality and a cognitive-domain label to each dataset. It reads only the evidence the deposit carries (BIDS description, README, participants, tasks, events, any DOI-linked abstract) and writes a label only above a per-category confidence threshold. Free text and event annotations stay human-curated; the model identifier, tagging time and taxonomy version are stored per dataset in \texttt{tagger\_meta} (Supplementary~S4).

\subsection*{Repair}\nrlabel{Repair}{sec:repair}

BIDS allows a fixed set of container formats per modality and \eegdash{} catalogues all of them, reading the signal through MNE-BIDS (the list is in Supplementary~S3; NWB, KRISS and ITAB are indexed from metadata until their readers land upstream). The repair layer has one function for each defect category that the audit (\nameref{sec:audit}) found to block automated loading of the OpenNeuro corpus. \emph{File-pointer} defects are broken internal references, typically a BrainVision header that names its marker or signal file under the wrong path. \emph{Metadata} defects are missing or invalid sidecar JSON and TSV encoding or value-type errors. \emph{Structural} defects are missing or malformed channels, events or participants tables, and montage or coordinate-system inconsistencies. \emph{Format-specific} defects are reader failures inside the signal file (EDF, EEGLAB, MEF3, FIF). Each function tests for one violation and applies the smallest correction that removes it, in memory; the deposit is never rewritten, so signal preservation holds by construction and is not measured separately.

\subsection*{ML-ready access}\nrlabel{ML-Ready Data Format}{sec:mlformat}

Each catalogued recording is held as an \texttt{EEGDashRaw} object that carries its record document and a storage pointer to the source archive. It can be filtered with no transfer; the signal bytes are fetched and cached when the \texttt{.raw} view is first accessed. On access, \eegdash{} converts the recording into the memory-mapped Braindecode format \citep{braindecode}, a subclass of the PyTorch \texttt{Dataset} \citep{paszke2019pytorch}. The recording becomes epochs or fixed-length windows; label vectors come from the BIDS event labels, and subject-level metadata (age, pathology, reaction time) is attached as annotations. Standard libraries \citep{paszke2019pytorch, pedregosa2011scikitlearn} consume the result without custom data-loading code; feature-based workflows use the table described next.

\subsection*{The feature module}\nrlabel{The feature module}{sec:features}

\begin{figure*}[tbp]
    \centering
    \includegraphics[width=\textwidth]{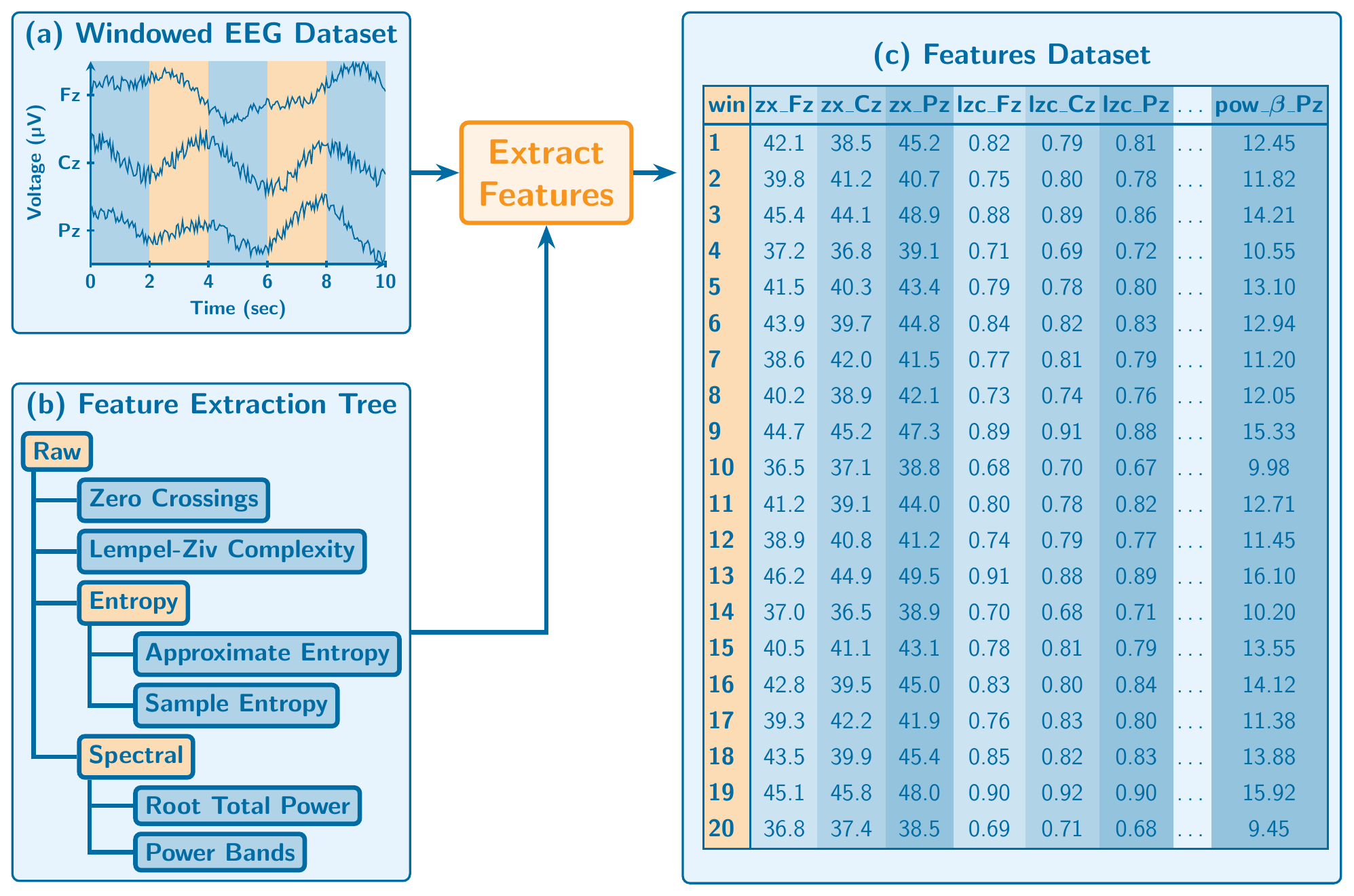}
    \caption{\textbf{A shared-computation tree turns windowed EEG into a features table in one call, reusing spectra across features.} (a)~Raw EEG from three channels is divided into 2~s windows. (b)~A feature extraction tree with 2 preprocessors and 6 features. (c)~Feature table (windows~$\times$~features) for 7 representative columns.}
    \figalt{Three-part workflow. Raw EEG traces are split into time windows, a feature extraction tree branches into zero crossings, entropy, and spectral features, and the output is a table with windows as rows and extracted feature values as columns.}
    \label{fig:features}
\end{figure*}

Classical models and inspectable diagnostics need hand-crafted features. The \texttt{eegdash.features} module turns windowed multichannel data into a windows~$\times$~features table through a feature-extraction tree in which features share intermediate steps (several spectral features reuse one power spectrum), batched over windows by PyTorch \citep{paszke2019pytorch} (Figure~\ref{fig:features}). The module ships 46 features and 10 preprocessors in six categories: signal statistics, spectral, complexity, dimensionality, connectivity and spatial (Supplementary~S5). Of the features, 34 are univariate and 11 bivariate; one, Common Spatial Patterns, is multivariate and trainable through a scikit-learn-style fit/transform interface \citep{pedregosa2011scikitlearn}. A feature set is extracted from a Braindecode dataset in one call; the result keeps the original labels and is usable as a PyTorch dataset or as a pandas dataframe. New features are plain Python functions with a few decorators; an inspect module lists what exists.

\subsection*{Loadability and compliance audit}\nrlabel{Loadability and compliance audit}{sec:audit}

The audit covers the OpenNeuro EEG corpus as catalogued in \eegdash{}~v0.6.0 (May~2026). Every recording was read through the canonical BIDS loader, \texttt{mne-bids}~0.19 on MNE-Python~1.13, and every dataset was run through the official BIDS validator, the NEMAR deployment of \texttt{@bids/validator}~2.4.1 on Deno~2.1.6. The validator completed on 548~datasets (10 timed out or failed to parse); the loader was run on 506~datasets and 114\,031~recordings; the 503~datasets with both results enter the compliance-versus-loadability comparison. NEMAR-hosted datasets are excluded: they are curated under \eegdash{} governance, all pass the validator and all load. The audited set predates the registry's cross-source de-duplication and still counts the Healthy Brain Network OpenNeuro deposits now recorded once through their NEMAR mirror. Per-record outcomes, error categories and validator outputs will be deposited on Zenodo upon acceptance. The registry counts reported in Results are a snapshot of \Nsnapshot{} (eegdash~\Nversion{}).

\subsection*{Ethics statement}

All datasets catalogued by \eegdash{} were published in public repositories with the ethical approval obtained by the original investigators. \eegdash{} collects no new human data and required no additional institutional review.

\section*{Results}\nrlabel{Results}{sec:results}

\subsection*{What the registry holds}\nrlabel{Archive scale and composition}{sec:records}

\begin{figure}[tbp]
\centering
\iflatexml
\includegraphics[width=0.481\textwidth]{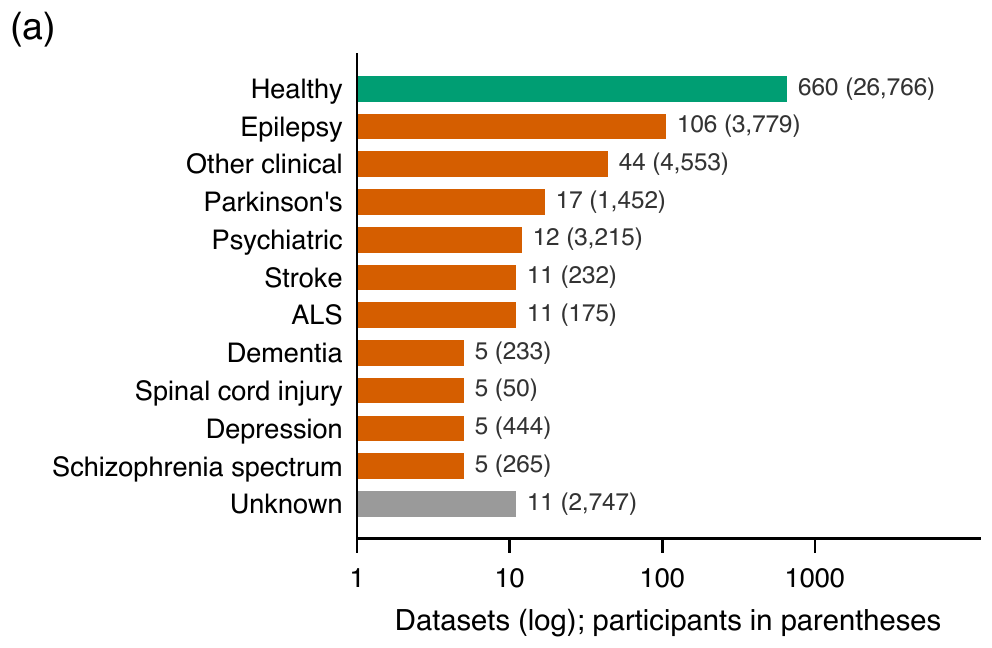}%
\includegraphics[width=0.499\textwidth]{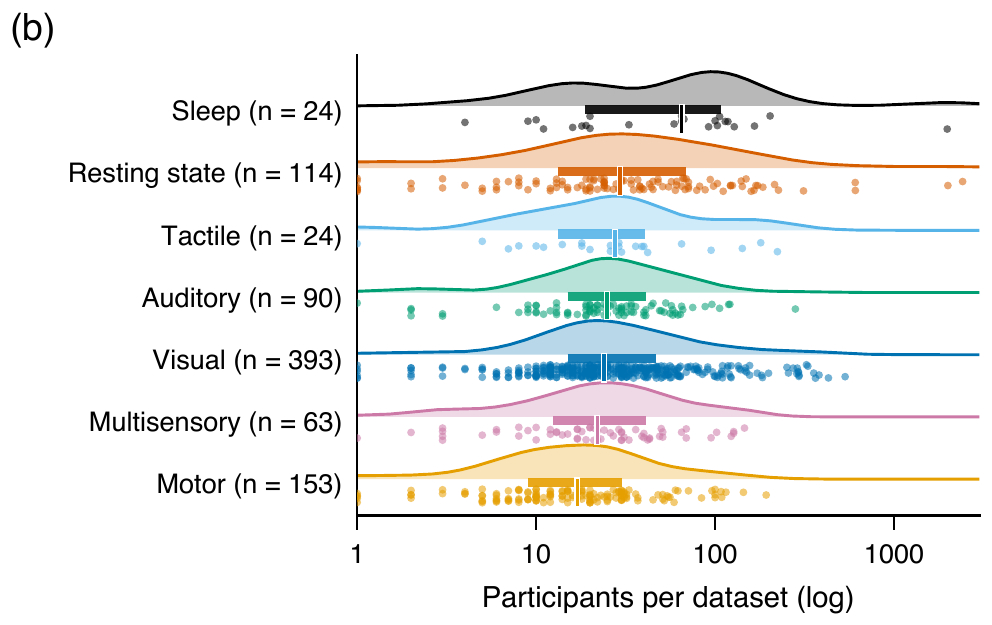}
\else
\includegraphics[width=0.476\textwidth]{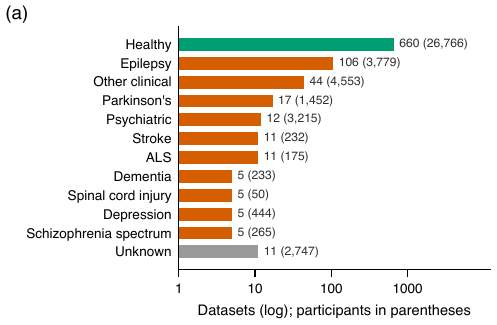}\hfill
\includegraphics[width=0.494\textwidth]{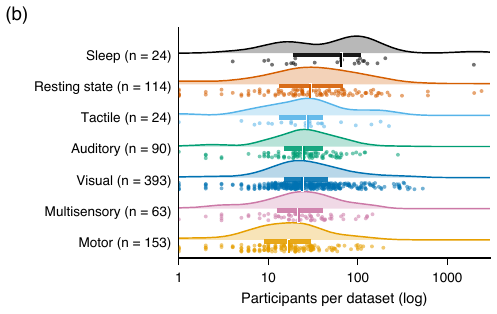}
\fi
\caption{\textbf{Who is in the registry} (\Nsnapshot{} snapshot). (a)~Datasets per population class, with the number of participants in parentheses; clinical classes with fewer than five datasets are pooled as ``Other clinical''. (b)~Participants per dataset by paradigm (density, individual datasets, interquartile box, median), with the paradigm derived from the tagger's experiment-type and stimulus-modality labels.}
\figalt{Two panels. a: horizontal bars of datasets per population class on a log scale, Healthy far largest, then Epilepsy, Other clinical, Parkinson's, Psychiatric, Stroke, ALS, Dementia, Spinal cord injury, Depression, Schizophrenia spectrum and Unknown. b: raincloud plots of participants per dataset on a log scale for Motor, Multisensory, Visual, Auditory, Tactile, Resting state and Sleep, medians between 17 and 65.}
\label{fig:composition}
\end{figure}

\eegdash{} catalogues \Ndatasets~datasets, \Nsubjects~participants, \Nrecords~recordings and \Nhours~hours across five modalities, each linked to its source deposit by a persistent identifier (Table~\ref{tab:archive}); the per-dataset catalogue is released as a supplementary data file. A dataset keeps its upstream BIDS layout and its declared BIDS version, so what a researcher fetches is the deposit unchanged. EEG accounts for 672 of the \Ndatasets~datasets and for most participants; EMG, with ten datasets, contributes a fifth of the hours through a few long continuous recordings. Healthy cohorts dominate and clinical populations form a minority tail (Figure~\ref{fig:composition}a); per-dataset cohorts are small and right-skewed across paradigms (Figure~\ref{fig:composition}b). These imbalances are inherited from what the community has shared, and cross-modality or cross-population analyses have to account for them.

\begin{table}[tbp]
\centering
\caption{\eegdash{} registry by recording modality (\Nsnapshot{} snapshot, eegdash~\Nversion{}). A multimodal dataset is counted in every modality row it contributes to, so Datasets, Subjects, Records and Hours do not sum to the deduplicated totals; Size is stored once under the primary modality and is additive. The EMG hours come from ten datasets of long continuous recordings (about 31~hours per participant).}
\label{tab:archive}
\begin{adjustbox}{max width=\columnwidth}
\begin{tabular}{@{}lrrrrr@{}}
\toprule
\textbf{Modality} & \textbf{Datasets$^{\dagger}$} & \textbf{Subjects$^{\dagger}$} & \textbf{Records$^{\dagger}$} & \textbf{Hours$^{\dagger}$} & \textbf{Size (TB)} \\
\midrule
EEG   & 672 &  37\,227 &  207\,445 &  113\,344 & 48.9 \\
iEEG  & 121 &   3\,773 &  18\,769 &   10\,640 &  9.6 \\
MEG   &  65 &   1\,517 &   11\,024 &    1\,244 & 11.4 \\
fNIRS &  31 &      866 &    8\,657 &       587 &  0.6 \\
EMG   &  10 &      769 &   34\,791 &   23\,903 &  1.3 \\
\midrule
\textbf{Total (unique)} & \textbf{\Ndatasets} & \textbf{\Nsubjects} & \textbf{\Nrecords} & \textbf{\Nhours} & \textbf{\Nsize} \\
\bottomrule
\end{tabular}
\end{adjustbox}
\par\smallskip
\footnotesize $^{\dagger}$Multimodal datasets are counted once per modality, so column sums exceed the deduplicated archive total.
\end{table}

\subsection*{The loader reads 95\,\% of OpenNeuro EEG recordings}\nrlabel{Loadability and compliance audit}{sec:validation}\nrlabel{Repair Coverage}{sec:repair-coverage}

The audit (\nameref{sec:audit}) answers the two questions left open in the Introduction for the OpenNeuro EEG corpus: loadability without per-dataset engineering, and whether validator compliance is the right gate for it. On 506~datasets and 114\,031~recordings, the canonical loader alone reads 108\,322~recordings (95.0\,\%): 411~datasets completely, 60 partially and 35 (6.9\,\%) not at all. Part of that baseline is \eegdash{}'s own work, since its generalisable fixes go upstream to \texttt{mne-bids} and raise the reading rate of every tool that uses it. The repair layer then handles the dataset-specific residual: re-running the failed loads through it returns 969 further recordings (109\,291; 95.8\,\%) and moves two datasets from partial to complete. The 4\,740~recordings in 35 unloadable and 58 partially loadable datasets remain, and are exposed as queryable metadata for tracking and upstream resolution.

Load failures fall into the four categories of the repair layer (\nameref{sec:repair}; functions listed in Supplementary~S3). Counting each dataset by its dominant category, metadata defects affect 55~datasets (9\,014~recordings), structural defects 40 (2\,921), file-pointer defects 37 (1\,870) and format-specific reader failures 27 (1\,929); 354~datasets load without repair. The counts in parentheses are all recordings in the affected datasets, not the failed ones. The categories sum to 513 because 7~empty or unresolved deposits with no loadable recording sit outside the 506-dataset denominator; the per-record tables keep the full distribution of secondary defects.

\subsection*{One in three OpenNeuro EEG datasets passes the BIDS validator}\nrlabel{Only One in Three OpenNeuro EEG Datasets Is BIDS-Compliant}{sec:validator-status}

\begin{figure*}[tbp]
    \centering
    \includegraphics[width=\textwidth]{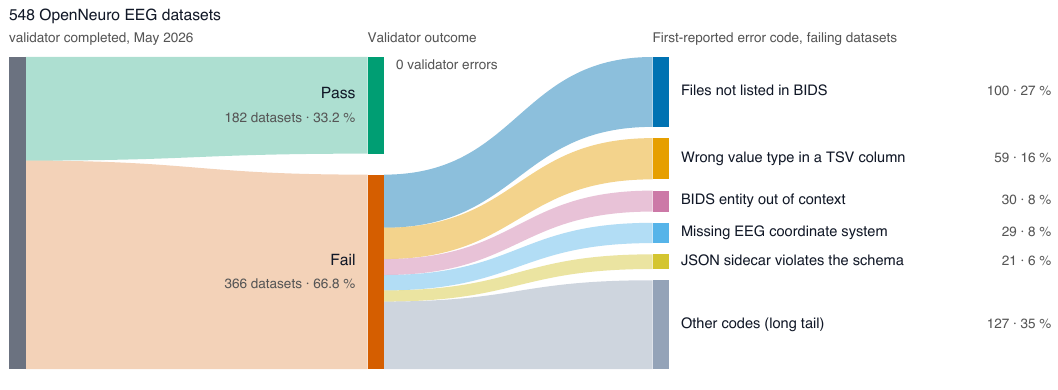}
    \caption{\textbf{One in three OpenNeuro EEG datasets passes the BIDS validator.} The 548~datasets on which the validator completed (May~2026 audit; 10 further datasets timed out or failed to parse), split by outcome and, for the 366 that fail, by their first-reported error code; the five most frequent codes are named and the rest pooled as ``Other''. Because each dataset is counted once, by its first-reported code, these counts differ from the per-code counts in the text and in Supplementary~S1, which count every dataset that raises a code.}
    \figalt{Flow diagram in three columns. Left: one block for the 548 OpenNeuro EEG datasets on which the validator completed. Middle: two blocks, pass with 182 datasets (33.2 percent) and fail with 366 datasets (66.8 percent). Right: the failing datasets split by first-reported error code into six blocks, files not listed in BIDS (100), wrong value type in a TSV column (59), BIDS entity out of context (30), missing EEG coordinate system (29), JSON sidecar violates the schema (21) and other codes (127).}
    \label{fig:validator-sankey}
\end{figure*}

A dataset is BIDS-compliant when the official validator reports zero errors. Of the 548~datasets on which the validator completed, 182 (33.2\,\%) pass; 366 (66.8\,\%) raise at least one error, 180\,697~errors and 3.41~million warnings in all (Figure~\ref{fig:validator-sankey}). Counting the 10~datasets on which the validator did not complete, 67.4\,\% of the audited corpus is not demonstrably compliant. OpenNeuro requires every dataset to pass the validator at upload \citep{markiewicz2021openneuro}, and the earlier survey of archives took that upload-time check as the compliance guarantee \citep{subash2023comparison}; the audit measures compliance as of May~2026, not at deposit. The two most frequent codes are wrong value types in tabular metadata (191~datasets) and files present on disk but not declared in BIDS (142); the next three are missing electrode coordinate systems (43), participant-identifier mismatches (37) and malformed HED tags (34) \citep{pernet2019eegbids}. The full code list, with the upstream condition each code traces to, is in Supplementary~S1.

\subsection*{Compliance does not predict loadability}\nrlabel{Compliance and Loadability}{sec:validator-vs-load}

On the 503~datasets with both results, the rank correlation between a dataset's validator-error count and the fraction of its recordings that load is near zero (Spearman $\rho = -0.05$; 95\,\% CI $[-0.14,\,0.04]$; $p \approx 0.26$); the rank coefficient is insensitive to the heavy skew of the error counts. Table~\ref{tab:validator-vs-load} crosses validator status with loader outcome and makes the compliance--loadability gap concrete. Of the 332~datasets that fail the validator, 219 load completely; of the 171 that pass, 37 do not load completely and 12 cannot be read at all. At this binary level the association is detectable ($\chi^2 = 8.3$, $df = 1$, $p \approx 0.004$) but small ($\phi = 0.13$): validator-clean datasets load completely in 78\,\% of cases, validator-failing ones in 66\,\%. A platform that admitted only validator-clean datasets would discard the 219 that load; one that admitted only datasets that load would carry 219 datasets with specification violations into every downstream analysis. \eegdash{} keeps both signals as queryable metadata on every dataset.

\begin{table}[tbp]
\centering
\caption{Validator status crossed with loader outcome for the 503~OpenNeuro EEG datasets with both results (dataset counts). Spearman correlation between validator-error count and the fraction of recordings that load: $\rho = -0.05$ (95\,\% CI $[-0.14,\,0.04]$, $p \approx 0.26$); on the table, $\chi^2 = 8.3$, $df = 1$, $p \approx 0.004$, $\phi = 0.13$.}
\label{tab:validator-vs-load}
\footnotesize
\begin{tabular}{@{}lrrr@{}}
\toprule
                          & \textbf{Loader 100\%} & \textbf{Loader partial/fail} & \textbf{total} \\
\midrule
\textbf{validator-broken} & 219                              & 113                                     & 332            \\
\textbf{validator-clean}  & 134                              &  37 (12 fully unreadable)               & 171            \\
\midrule
\textbf{total}            & 353                              & 150                                     & 503            \\
\bottomrule
\end{tabular}
\end{table}

\subsection*{Every catalogued dataset reaches a PyTorch DataLoader}\nrlabel{End-to-End ML Loadability}{sec:ml-verification}\nrlabel{Interface evaluation}{sec:ux-eval}

A catalogued dataset goes from import to a training loop in three lines: the class is imported, wrapped in a standard \texttt{DataLoader} and iterated, and repair, windowing and caching run lazily on first access (Listing~\ref{lst:three-line}). The tutorial gallery at \url{https://eegdash.org} publishes end-to-end notebooks for P300 decoding, eyes-open versus eyes-closed classification, age regression, sex classification and transfer learning; each loads data through the \texttt{eegdash} API, preprocesses with Braindecode and trains a model with no custom adapter. The exercise is structural, not accuracy-driven: it shows that a public dataset class reaches a working training loop through the published API without custom data-loading code.

\begin{lstlisting}[caption={Three-line training-ready pipeline using the \eegdash{} registry, a standard PyTorch \texttt{DataLoader}, and lazy repair, windowing and caching.}, label=lst:three-line]
# requires: eegdash>=0.9.1, torch>=2.0
# 1. Pick any catalogued dataset
from eegdash.dataset import DS001849
# 2. Wrap in a standard PyTorch loader
from torch.utils.data import DataLoader
loader = DataLoader(DS001849(cache_dir="./data"),
                    batch_size=32)
# 3. Iterate: repair, windowing and caching are all lazy
for X, y, _ in loader: ...
\end{lstlisting}

Three properties of the interface hold by construction and are checked rather than measured (Supplementary~S7) \citep{akhtar2024croissant}. Consistency: a continuous-integration test asserts that every generated class returns the same \texttt{EEGDashDataset} contract. Completeness: every tutorial in the gallery issues its discovery query through the public API alone. Conciseness: loading takes under ten lines and an end-to-end example under eighty. Perceived readability and understandability would need user ratings and remain open.

\section*{Discussion}\nrlabel{Discussion}{sec:discussion}

\eegdash{} does three things. It indexes the signal metadata of \Ndatasets{} public datasets, so a cohort is chosen before any signal is downloaded. It loads each of them into a PyTorch dataset through one class, with repair, windowing and caching applied lazily, and recovers 95.8\,\% of OpenNeuro EEG recordings; a catalogued dataset trains in three lines (Listing~\ref{lst:three-line}). And it measured validator compliance against loader success on 503 OpenNeuro EEG datasets and found no relation between them ($\rho = -0.05$): 219 datasets fail the validator and load, 37 pass and do not.

Table~\ref{tab:comparison} places \eegdash{} among the platforms that serve neural data. Archives host at scale and stop at storage: OpenNeuro, NEMAR and DANDI deliver raw files \citep{markiewicz2021openneuro, delorme2022nemar}, and the Canadian Open Neuroscience Platform federates datasets and checks that they download, not that they load \citep{poline2023conp}. Curated libraries ship loaders over a hand-picked set \citep{chevallier2024moabb, zhang2024torcheeg}. Two neighbouring communities have built the layer in between in their own way. The NWB ecosystem standardises 47 acquisition formats before deposit through NeuroConv \citep{rubel2022nwb, mayorquin2025neuroconv}; the MRI community, on the same OpenNeuro and BIDS foundation, turns heterogeneous deposits into analysis-ready derivatives with BIDS Apps and fMRIPrep \citep{gorgolewski2017bidsapps, esteban2019fmriprep}. For neural time series the obstacle is loadability and format drift, not spatial normalisation, and the FAIR principles, which BIDS compliance satisfies at the level of metadata standards \citep{wilkinson2016fair}, say nothing about whether a loader can parse the signal. \eegdash{} fills that content-level layer and connects it outward: datasets are exported to Hugging Face with Croissant metadata \citep{lhoest2021datasets, akhtar2024croissant}, consumed by NeuralSet, NeuralBench and MOABB \citep{king2026neuralset, banville2026neuralbench, chevallier2024moabb}, dispatched to HPC through the Neuroscience Gateway, and the EEG2025 Foundation Challenge runs on a dedicated dataset class.

The compliance--loadability gap has a mechanical explanation. The validator and the loader are both living research software. The validator's rule set tightens with each release, so a dataset that passed under one version can raise errors under the next \citep{dicosmo2026code, lamprecht2020fair4rse}; OpenNeuro lets uploaders change files after a snapshot without re-validation \citep{markiewicz2021openneuro}; and the loader changes with each \texttt{mne-bids} release. A dataset's status on either axis therefore drifts with the toolchain, which is why the audit pins versions and releases per-record outcomes, so that loadability can be re-measured as a reproducibility step \citep{kapoor2023leakage, niso2022reproducibleeeg}. The same mechanism decides where a fix belongs: a correction sent to \texttt{mne-bids} raises the reading rate of every tool that uses it, whereas a repair kept inside one platform fixes one copy of the corpus.

The claims have bounds. The audit covers the OpenNeuro EEG corpus under pinned loader and validator versions and reports associations, not causes; NEMAR deposits and the other four modalities were not audited. HED event annotation is partial, so cross-study queries by task are valid on the annotated subset only, and the tagger's labels have not yet been checked against a human-labelled sample. EEG dominates the registry and healthy participants outnumber any clinical population (Table~\ref{tab:archive}, Figure~\ref{fig:composition}a), so cross-modality and clinical conclusions have to account for the imbalance. The repair layer corrects format defects only; acquisition artefacts are left to the user, and a derivatives layer built on EEGPrep (\url{https://eegprep.org}) with standardised quality metrics is the next step. \eegdash{} is infrastructure for methods research and benchmarking, not for clinical or diagnostic use.

\eegdash{} carries validator status and loader outcome as two queryable signals on every dataset, so the measurement is part of the archive and not of a single audit. Readiness for machine learning is a property of an archive that has to be measured; compliance with the deposit standard is not a proxy for it.

\begin{table*}[tbp]
\centering
\caption{\eegdash{} beside nine neural-data platforms, grouped by role. MEEG: EEG, iEEG, MEG, fNIRS and EMG. The \eegdash{} row is the \Nsnapshot{} snapshot; the other rows follow each platform's public documentation as of May~2026. Symbols are defined below the table.}
\label{tab:comparison}
% symbols: native / partial or via adapter / not provided / none or not applicable
\newcommand{\cyes}{\tikz[baseline=-0.55ex]\fill[eegblue] (0,0) circle (0.72ex);}
\newcommand{\chalf}{\tikz[baseline=-0.55ex]{\draw[eegblue,line width=0.5pt] (0,0) circle (0.72ex);\fill[eegblue] (0,0.72ex) arc (90:270:0.72ex) -- cycle;}}
\newcommand{\cno}{\tikz[baseline=-0.55ex]\draw[eeggray!55,line width=0.5pt] (0,0) circle (0.72ex);}
\newcommand{\cna}{{\color{eeggray!80}--}}
\newcommand{\tgroup}[1]{\multicolumn{10}{@{}l}{\rule{0pt}{2.4ex}\textit{\color{eeggray!95!black}#1}}\\}
\setlength{\tabcolsep}{5pt}
\renewcommand{\arraystretch}{1.12}
\begin{adjustbox}{max width=\textwidth}
\small
\begin{tabular}{@{}l r l @{\hspace{9pt}} c c c l @{\hspace{9pt}} c c l @{}}
\toprule
 & \multicolumn{2}{c}{\textbf{Scale}} & \multicolumn{4}{c}{\textbf{ML readiness}} & \multicolumn{3}{c}{\textbf{Access}} \\
\cline{2-3}\cline{4-7}\cline{8-10}
\textbf{Platform} & \textbf{Datasets} & \textbf{Modalities} & \textbf{ML-ready} & \textbf{BIDS repair} & \textbf{Feature bank} & \textbf{Events} & \textbf{Python API} & \textbf{DataLoader} & \textbf{HPC} \\
\midrule
\tgroup{Raw BIDS / NWB archives}
\rowcolor{eegblue!9}\textbf{\eegdash{}} & \textbf{\Ndatasets} & MEEG (5) & \cyes & \cyes & \cyes & BIDS & \cyes & \cyes & NSG \\
OpenNeuro & 1\,627 & multi (6) & \cno & \cno & \cno & BIDS* & \chalf & \chalf & Brainlife \\
NEMAR & 358 & EEG, MEG & \cno & \cno & \cno & BIDS* & \cyes & \chalf & NSG \\
DANDI & 1\,074 & multi (NWB) & \cno & \cna & \cno & NWB & \chalf & \chalf & \cna \\
PhysioNet~\citep{goldberger2000physionet} & 100+ & multi & \cno & \cna & \cno & varies & \chalf & \chalf & \cna \\
TUH EEG~\citep{obeid2016tuh} & 1 corpus & EEG (clinical) & \cno & \cna & \cno & clinical & \chalf & \chalf & \cna \\
\tgroup{Curated ML libraries}
MOABB & 157 & EEG & \chalf$^{\dagger}$ & \cna & \cno & paradigm & \cyes & \chalf$^{\dagger}$ & \cna \\
Brainsets~\citep{azabou2023unified} & 9 & spikes, mixed & \cyes & \cna & \cyes & custom & \cyes & \cyes & \cna \\
TorchEEG & 5+ & EEG & \cyes & \cna & \cyes & custom & \cyes & \cyes & \cna \\
\tgroup{Multi-modal orchestrator}
NeuralSet & 100+ (framework) & multi (7) & \cyes & \cna & \chalf & BIDS/custom & \cyes & \cyes & SLURM \\
\bottomrule
\end{tabular}
\end{adjustbox}
\par\smallskip
\parbox{\textwidth}{\footnotesize\raggedright \cyes{} native \quad \chalf{} partial, or through a third-party adapter, CLI or web API \quad \cno{} not provided \quad \cna{} none or not applicable. \quad *Event annotation partial in the source datasets and not enforced at upload. \quad $^{\dagger}$NumPy arrays for scikit-learn; PyTorch through Braindecode.}
\end{table*}

\section*{Conclusion}

\eegdash{} turns the public BIDS archive into data a model can train on: a registry of \Ndatasets{} datasets queried before any download, one class per dataset that loads with lazy repair, and an audit showing that BIDS compliance and loadability are different properties and have to be tracked separately. Everything described here is at \url{https://eegdash.org}: the registry, the browser and the package, with installation, filtering and HPC dispatch documented in Supplementary~S2. The audit tables will be deposited on Zenodo upon acceptance.

\section*{Supplementary Data}
Supplementary data are available as ancillary files with this arXiv preprint: S1, the full validator code table with the upstream condition each code traces to; S2, installation, filtering and HPC dispatch; S3, the container formats and the repair taxonomy; S4, the tagging mechanics; S5, the feature bank and its decorator API; S6, the dataset catalogue, released as the data file \texttt{dataset\_catalogue.csv} (\Ndatasets{} rows); S7, the interface checks. The audit tables (per-record loader outcomes, repair categories and validator outputs for the OpenNeuro EEG corpus) will be deposited on Zenodo upon acceptance.

\section*{Funding}
This work was supported by the National Institutes of Health [grant numbers R24MH120037, U24EB029005]; and the National Science Foundation [grant numbers 1935749, 2423943]. The content is solely the responsibility of the authors and does not necessarily represent the official views of the funding agencies.

\section*{Acknowledgements}
We thank Hubert Banville for his insightful discussions and continued feedback on the first version of the manuscript. We are grateful to Alexandre Gramfort, Thomas Moreau, Eric Larson, Daniel McCloy, and Scott Huberty, and to the broader open-source neuroscience community, for the tools and standards on which this work builds. Finally, we thank the authors and participants of the datasets catalogued in \eegdash{}, whose openly shared recordings make this resource possible.

\section*{Author Contributions}
B.A.\ contributed to conceptualisation, methodology, software, data curation, visualisation, writing (original draft), and writing (review \& editing).
A.Do.\ contributed to conceptualisation, methodology, software, visualisation, writing \& review.
P.G.\ contributed to methodology, software, and writing (review \& editing).
A.M.\ contributed to resources and infrastructure (NSG integration).
G.A.\ contributed to methodology, software and review.
D.T.\ contributed to software, data curation, and infrastructure.
K.K., A.J.\ contributed to software and data curation.
O.S.\ contributed to conceptualisation, supervision and funding acquisition.
A.De.\ contributed to conceptualisation, supervision, funding acquisition, and writing (review \& editing).

\section*{Conflict of Interest}
B.A.\ has been an employee of Yneuro since November~2025. The remaining authors declare no competing interests.

\section*{Data Availability}
All \Ndatasets{} datasets catalogued in \eegdash{} (v\Nversion{}) are accessible through the web browser at \url{https://eegdash.org} and the API at \url{https://data.eegdash.org}, without login or registration; the resource will remain available for at least two years after publication. Each dataset is linked to its source deposit (OpenNeuro, NEMAR) by a persistent identifier and keeps the licence assigned there; the per-dataset licence is recorded in the catalogue (\texttt{dataset\_catalogue.csv}, Supplementary~S6; browsable at \url{https://eegdash.org/dataset_summary.html} and shipped with the package as \texttt{dataset\_summary.csv}) and in the Croissant~1.0 descriptors, and users should cite both this article and the source datasets, whose accessions are listed with each entry. Long-term preservation of the recordings is provided by the source repositories.

\section*{Code Availability}
The \eegdash{} Python package (v\Nversion{}) is released under the BSD-3-Clause licence on GitHub (\url{https://github.com/eegdash/EEGDash}; \texttt{pip install eegdash}), with documentation, tutorials and per-dataset API pages at \url{https://eegdash.org}, and archived on Zenodo at \href{https://doi.org/10.5281/zenodo.20646753}{10.5281/zenodo.20646753} (concept DOI, resolving to the latest version). Braindecode \citep{braindecode}, the deep-learning dependency, is a separate open-source package.

\bibliographystyle{database}
\bibliography{references}

\begin{thebibliography}{10}

\bibitem[{gorgolewski2016bids(2016)}]{gorgolewski2016bids}
Gorgolewski KJ, Auer T, Calhoun VD et~al.
\newblock The brain imaging data structure, a format for organizing and
  describing outputs of neuroimaging experiments.
\newblock Sci Data 2016;3:160044.
\newblock \href {https://doi.org/10.1038/sdata.2016.44}
  {\nolinkurl{https://doi.org/10.1038/sdata.2016.44}}.

\bibitem[{pernet2019eegbids(2019)}]{pernet2019eegbids}
Pernet CR, Appelhoff S, Gorgolewski KJ et~al.
\newblock {EEG-BIDS}, an extension to the brain imaging data structure for
  electroencephalography.
\newblock Sci Data 2019;6:103.
\newblock \href {https://doi.org/10.1038/s41597-019-0104-8}
  {\nolinkurl{https://doi.org/10.1038/s41597-019-0104-8}}.

\bibitem[{markiewicz2021openneuro(2021)}]{markiewicz2021openneuro}
Markiewicz CJ, Gorgolewski KJ, Feingold F et~al.
\newblock The {OpenNeuro} resource for sharing of neuroscience data.
\newblock eLife 2021;10:e71774.
\newblock \href {https://doi.org/10.7554/eLife.71774}
  {\nolinkurl{https://doi.org/10.7554/eLife.71774}}.

\bibitem[{delorme2022nemar(2022)}]{delorme2022nemar}
Delorme A, Truong D, Youn C et~al.
\newblock {NEMAR}: an open access data, tools and compute resource operating on
  neuroelectromagnetic data.
\newblock Database (Oxford) 2022;2022:baac096.
\newblock \href {https://doi.org/10.1093/database/baac096}
  {\nolinkurl{https://doi.org/10.1093/database/baac096}}.

\bibitem[{wilkinson2016fair(2016)}]{wilkinson2016fair}
Wilkinson MD, Dumontier M, Aalbersberg IJ et~al.
\newblock The {FAIR} guiding principles for scientific data management and
  stewardship.
\newblock Sci Data 2016;3:160018.
\newblock \href {https://doi.org/10.1038/sdata.2016.18}
  {\nolinkurl{https://doi.org/10.1038/sdata.2016.18}}.

\bibitem[{kostas2021bendr(2021)}]{kostas2021bendr}
Kostas D, Aroca-Ouellette S, Rudzicz F.
\newblock {BENDR}: Using Transformers and a Contrastive Self-Supervised
  Learning Task to Learn from Massive Amounts of {EEG} Data.
\newblock Front Hum Neurosci 2021;15:653659.
\newblock \href {https://doi.org/10.3389/fnhum.2021.653659}
  {\nolinkurl{https://doi.org/10.3389/fnhum.2021.653659}}.

\bibitem[{yang2023biot(2023)}]{yang2023biot}
Yang C, Westover MB, Sun J.
\newblock {BIOT}: Biosignal Transformer for Cross-data Learning in the Wild.
\newblock In: Advances in Neural Information Processing Systems (NeurIPS).
  vol.~36. New Orleans, LA, USA; 2023. p. 78240-60.
\newblock \href {https://doi.org/10.52202/075280-3420}
  {\nolinkurl{https://doi.org/10.52202/075280-3420}}.

\bibitem[{jiang2024labram(2024)}]{jiang2024labram}
Jiang WB, Zhao LM, Lu BL.
\newblock Large Brain Model for Learning Generic Representations with
  Tremendous {EEG} Data in {BCI}.
\newblock In: International Conference on Learning Representations (ICLR).
  Vienna, Austria; 2024. \href {https://arxiv.org/abs/2405.18765}
  {\nolinkurl{arXiv:2405.18765}}.

\bibitem[{xiao2025brainomni(2025)}]{xiao2025brainomni}
Xiao Q, Cui Z, Zhang C et~al.
\newblock {BrainOmni}: A Brain Foundation Model for Unified {EEG} and {MEG}
  Signals.
\newblock In: Advances in Neural Information Processing Systems (NeurIPS).
  vol.~38. San Diego, CA, USA; 2025. \href
  {https://doi.org/10.52202/085713-1375}
  {\nolinkurl{https://doi.org/10.52202/085713-1375}}.

\bibitem[{sculley2015debt(2015)}]{sculley2015debt}
Sculley D, Holt G, Golovin D et~al.
\newblock Hidden Technical Debt in Machine Learning Systems.
\newblock In: Advances in Neural Information Processing Systems (NeurIPS).
  vol.~28. Montreal, Canada; 2015. p. 2503-11.

\bibitem[{chevallier2024moabb(2024)}]{chevallier2024moabb}
Chevallier S, Carrara I, Aristimunha B et~al.
\newblock The largest {EEG}-based {BCI} reproducibility study for open science:
  the {MOABB} benchmark; 2024.
\newblock \href {https://arxiv.org/abs/2404.15319}
  {\nolinkurl{arXiv:2404.15319}}.

\bibitem[{zhang2024torcheeg(2024)}]{zhang2024torcheeg}
Zhang Z, Zhong SH, Liu Y.
\newblock {TorchEEGEMO}: A deep learning toolbox towards {EEG}-based emotion
  recognition.
\newblock Expert Syst Appl 2024;249:123550.
\newblock \href {https://doi.org/10.1016/j.eswa.2024.123550}
  {\nolinkurl{https://doi.org/10.1016/j.eswa.2024.123550}}.

\bibitem[{westner2025cycling(2025)}]{westner2025cycling}
Westner BU, McCloy DR, Larson E et~al.
\newblock Cycling on the Freeway: The perilous state of open-source
  neuroscience software.
\newblock Imaging Neurosci 2025;3:imag\_a\_00554.
\newblock \href {https://doi.org/10.1162/imag_a_00554}
  {\nolinkurl{https://doi.org/10.1162/imag_a_00554}}.

\bibitem[{subash2023comparison(2023)}]{subash2023comparison}
Subash P, Gray A, Boswell M et~al.
\newblock A comparison of neuroelectrophysiology databases.
\newblock Sci Data 2023;10:719.
\newblock \href {https://doi.org/10.1038/s41597-023-02614-0}
  {\nolinkurl{https://doi.org/10.1038/s41597-023-02614-0}}.

\bibitem[{mohan2021datastalls(2021)}]{mohan2021datastalls}
Mohan J, Phanishayee A, Raniwala A et~al.
\newblock Analyzing and Mitigating Data Stalls in {DNN} Training.
\newblock Proc VLDB Endow 2021;14:771-84.
\newblock \href {https://doi.org/10.14778/3446095.3446100}
  {\nolinkurl{https://doi.org/10.14778/3446095.3446100}}.

\bibitem[{wagner2022fairlybig(2022)}]{wagner2022fairlybig}
Wagner AS, Waite LK, Wierzba M et~al.
\newblock {FAIRly} Big: A Framework for Computationally Reproducible Processing
  of Large-Scale Data.
\newblock Sci Data 2022;9:80.
\newblock \href {https://doi.org/10.1038/s41597-022-01163-2}
  {\nolinkurl{https://doi.org/10.1038/s41597-022-01163-2}}.

\bibitem[{gramfort2013mne(2013)}]{gramfort2013mne}
Gramfort A, Luessi M, Larson E et~al.
\newblock {MEG} and {EEG} data analysis with {MNE-Python}.
\newblock Front Neurosci 2013;7:267.
\newblock \href {https://doi.org/10.3389/fnins.2013.00267}
  {\nolinkurl{https://doi.org/10.3389/fnins.2013.00267}}.

\bibitem[{appelhoff2019mnebids(2019)}]{appelhoff2019mnebids}
Appelhoff S, Sanderson M, Brooks T et~al.
\newblock {MNE-BIDS}: Organizing electrophysiological data into the {BIDS}
  format and facilitating their analysis.
\newblock J Open Source Softw 2019;4:1896.
\newblock \href {https://doi.org/10.21105/joss.01896}
  {\nolinkurl{https://doi.org/10.21105/joss.01896}}.

\bibitem[{braindecode(2025)}]{braindecode}
Aristimunha B, Guetschel P, Wimpff M et~al.
\newblock Braindecode: toolbox for decoding raw electrophysiological brain data
  with deep learning models. Zenodo; 2025.
\newblock \href {https://doi.org/10.5281/zenodo.17699192}
  {\nolinkurl{https://doi.org/10.5281/zenodo.17699192}}.

\bibitem[{paszke2019pytorch(2019)}]{paszke2019pytorch}
Paszke A, Gross S, Massa F et~al.
\newblock {PyTorch}: An imperative style, high-performance deep learning
  library.
\newblock In: Advances in Neural Information Processing Systems. vol.~32.
  Vancouver, Canada: Curran Associates, Inc.; 2019. p. 8026-37.

\bibitem[{sivagnanam2013nsg(2013)}]{sivagnanam2013nsg}
Sivagnanam S, Astakhov V, Yoshimoto K et~al.
\newblock A neuroscience gateway.
\newblock In: Proceedings of the Conference on Extreme Science and Engineering
  Discovery Environment: Gateway to Discovery. San Diego, CA, USA: ACM; 2013.
  \href {https://doi.org/10.1145/2484762.2484816}
  {\nolinkurl{https://doi.org/10.1145/2484762.2484816}}.

\bibitem[{robbins2021hed(2021)}]{robbins2021hed}
Robbins K, Truong D, Appelhoff S et~al.
\newblock Capturing the nature of events and event context using {Hierarchical
  Event Descriptors} ({HED}).
\newblock Neuroimage 2021;245:118766.
\newblock \href {https://doi.org/10.1016/j.neuroimage.2021.118766}
  {\nolinkurl{https://doi.org/10.1016/j.neuroimage.2021.118766}}.

\bibitem[{pedregosa2011scikitlearn(2011)}]{pedregosa2011scikitlearn}
Pedregosa F, Varoquaux G, Gramfort A et~al.
\newblock Scikit-learn: Machine learning in {Python}.
\newblock J Mach Learn Res 2011;12:2825-30.

\bibitem[{akhtar2024croissant(2024)}]{akhtar2024croissant}
Akhtar M, Benjelloun O, Conforti C et~al.
\newblock Croissant: A Metadata Format for {ML}-Ready Datasets.
\newblock In: Advances in Neural Information Processing Systems (NeurIPS),
  Datasets and Benchmarks Track. vol.~37. Vancouver, Canada; 2024. p. 82133-48.
\newblock \href {https://doi.org/10.52202/079017-2610}
  {\nolinkurl{https://doi.org/10.52202/079017-2610}}.

\bibitem[{poline2023conp(2023)}]{poline2023conp}
Poline JB, Das S, Glatard T et~al.
\newblock Data and Tools Integration in the Canadian Open Neuroscience
  Platform.
\newblock Sci Data 2023;10:189.
\newblock \href {https://doi.org/10.1038/s41597-023-01946-1}
  {\nolinkurl{https://doi.org/10.1038/s41597-023-01946-1}}.

\bibitem[{rubel2022nwb(2022)}]{rubel2022nwb}
R{\"u}bel O, Tritt A, Ly R et~al.
\newblock The {Neurodata Without Borders} ecosystem for neurophysiological data
  science.
\newblock eLife 2022;11:e78362.
\newblock \href {https://doi.org/10.7554/eLife.78362}
  {\nolinkurl{https://doi.org/10.7554/eLife.78362}}.

\bibitem[{mayorquin2025neuroconv(2025)}]{mayorquin2025neuroconv}
Mayorquin H, Baker C, Adkisson-Floro P et~al.
\newblock {NeuroConv}: Streamlining Neurophysiology Data Conversion to the
  {NWB} Standard.
\newblock In: Proceedings of the 24th {Python} in Science Conference ({SciPy}).
  Tacoma, WA, USA; 2025. p. 245-58.
\newblock \href {https://doi.org/10.25080/cehj4257}
  {\nolinkurl{https://doi.org/10.25080/cehj4257}}.

\bibitem[{gorgolewski2017bidsapps(2017)}]{gorgolewski2017bidsapps}
Gorgolewski KJ, Alfaro-Almagro F, Auer T et~al.
\newblock {BIDS} apps: Improving ease of use, accessibility, and
  reproducibility of neuroimaging data analysis methods.
\newblock PLoS Comput Biol 2017;13:e1005209.
\newblock \href {https://doi.org/10.1371/journal.pcbi.1005209}
  {\nolinkurl{https://doi.org/10.1371/journal.pcbi.1005209}}.

\bibitem[{esteban2019fmriprep(2019)}]{esteban2019fmriprep}
Esteban O, Markiewicz CJ, Blair RW et~al.
\newblock {fMRIPrep}: a robust preprocessing pipeline for functional {MRI}.
\newblock Nat Methods 2019;16:111-6.
\newblock \href {https://doi.org/10.1038/s41592-018-0235-4}
  {\nolinkurl{https://doi.org/10.1038/s41592-018-0235-4}}.

\bibitem[{lhoest2021datasets(2021)}]{lhoest2021datasets}
Lhoest Q, Villanova~del Moral A, Jernite Y et~al.
\newblock Datasets: A Community Library for Natural Language Processing.
\newblock In: Proceedings of the 2021 Conference on Empirical Methods in
  Natural Language Processing: System Demonstrations. Punta Cana, Dominican
  Republic; 2021. p. 175-84.
\newblock \href {https://doi.org/10.18653/v1/2021.emnlp-demo.21}
  {\nolinkurl{https://doi.org/10.18653/v1/2021.emnlp-demo.21}}.

\bibitem[{king2026neuralset(2026)}]{king2026neuralset}
King JR, Bel C, Evanson L et~al.
\newblock NeuralSet: A High-Performing Python Package for Neuro-AI; 2026.
\newblock \href {https://arxiv.org/abs/2605.03169}
  {\nolinkurl{arXiv:2605.03169}}.

\bibitem[{banville2026neuralbench(2026)}]{banville2026neuralbench}
Banville H, d'Ascoli S, Dahan S et~al.
\newblock {NeuralBench}: A Unifying Framework to Benchmark NeuroAI Models;
  2026.
\newblock \href {https://arxiv.org/abs/2605.08495}
  {\nolinkurl{arXiv:2605.08495}}.

\bibitem[{dicosmo2026code(2026)}]{dicosmo2026code}
Di~Cosmo R, Granger S, Hinsen K et~al.
\newblock {CODE} beyond {FAIR}: a roadmap for reusable research software.
\newblock Sci Data 2026;13:514.
\newblock \href {https://doi.org/10.1038/s41597-026-06705-6}
  {\nolinkurl{https://doi.org/10.1038/s41597-026-06705-6}}.

\bibitem[{lamprecht2020fair4rse(2020)}]{lamprecht2020fair4rse}
Lamprecht AL, Garcia L, Kuzak M et~al.
\newblock Towards {FAIR} principles for research software.
\newblock Data Sci 2020;3:37-59.
\newblock \href {https://doi.org/10.3233/DS-190026}
  {\nolinkurl{https://doi.org/10.3233/DS-190026}}.

\bibitem[{kapoor2023leakage(2023)}]{kapoor2023leakage}
Kapoor S, Narayanan A.
\newblock Leakage and the reproducibility crisis in machine-learning-based
  science.
\newblock Patterns (N Y) 2023;4:100804.
\newblock \href {https://doi.org/10.1016/j.patter.2023.100804}
  {\nolinkurl{https://doi.org/10.1016/j.patter.2023.100804}}.

\bibitem[{niso2022reproducibleeeg(2021)}]{niso2022reproducibleeeg}
Niso G, Krol LR, Combrisson E et~al.
\newblock Enhancing reproducibility in developmental {EEG} research: {BIDS},
  cluster-based permutation tests, and effect sizes.
\newblock Dev Cogn Neurosci 2021;52:101036.
\newblock \href {https://doi.org/10.1016/j.dcn.2021.101036}
  {\nolinkurl{https://doi.org/10.1016/j.dcn.2021.101036}}.

\bibitem[{goldberger2000physionet(2000)}]{goldberger2000physionet}
Goldberger AL, Amaral LAN, Glass L et~al.
\newblock {PhysioBank}, {PhysioToolkit}, and {PhysioNet}: Components of a New
  Research Resource for Complex Physiologic Signals.
\newblock Circulation 2000;101:e215-20.
\newblock \href {https://doi.org/10.1161/01.cir.101.23.e215}
  {\nolinkurl{https://doi.org/10.1161/01.cir.101.23.e215}}.

\bibitem[{obeid2016tuh(2016)}]{obeid2016tuh}
Obeid I, Picone J.
\newblock The Temple University Hospital {EEG} Data Corpus.
\newblock Front Neurosci 2016;10:196.
\newblock \href {https://doi.org/10.3389/fnins.2016.00196}
  {\nolinkurl{https://doi.org/10.3389/fnins.2016.00196}}.

\bibitem[{azabou2023unified(2023)}]{azabou2023unified}
Azabou M, Arora V, Ganesh V et~al.
\newblock A Unified, Scalable Framework for Neural Population Decoding.
\newblock In: Advances in Neural Information Processing Systems. vol.~36; 2023.
  Citation requested by the \texttt{brainsets} package,
  \url{https://github.com/neuro-galaxy/brainsets}.

\end{thebibliography}

\end{document}